\documentclass{ws-ijmpe}

\newcommand{\AP}[3]{Ann.\ Phys.\ {\bf #1},\ #2 (#3)}

\newcommand{\NPA}[3]{Nucl.\ Phys.\ {\bf A#1},\ #2 (#3)}

\newcommand{\PAN}[3]{Phys.\ Atom.\ Nucl.\ {\bf #1},\ #2 (#3)}

\newcommand{\PLB}[3]{Phys.\ Lett.\ {\bf B#1},\ #2 (#3)}

\newcommand{\PRL}[3]{Phys.\ Rev.\ Lett.\ {\bf #1},\ #2 (#3)}
\newcommand{\PRR}[3]{Phys.\ Rev.\ {\bf #1},\ #2 (#3)}

\newcommand{\PRD}[3]{Phys.\ Rev.\ {\bf D#1},\ #2 (#3)}

\renewcommand\a{\alpha}
\renewcommand\b{\beta}
\renewcommand\d{\delta}
\renewcommand\k{\kappa}
\renewcommand\l{\lambda}

\renewcommand\t{\tau}

\renewcommand\c{\chi}
\renewcommand\j{\psi}

\newcommand\e{\epsilon}
\newcommand\g{\gamma}

\newcommand\m{\mu}
\newcommand\n{\nu}

\newcommand\p{\pi}
\newcommand\h{\theta}
\newcommand\s{\sigma}
\newcommand\f{\phi}

\newcommand\ve{\varepsilon}

\renewcommand\L{\Lambda}
\renewcommand\P{\Pi}

\renewcommand\O{\Omega}

\newcommand\D{\Delta}

\newcommand{\eq}[1]{Eq.\ (\ref{#1})}

\newcommand\lb{\left(}
\newcommand\rb{\right)}

\newcommand{\lan}{\langle}
\newcommand{\ran}{\rangle}

\newcommand\ra{\rightarrow}

\newcommand{\non}{\nonumber\\}
\newcommand\pt{\partial}
\newcommand{\mf}{{\rm mf}}
\newcommand{\Tr}{{\rm Tr}}

\newcommand{\im}{{\rm{Im}}}
\newcommand{\cl}{{\cal L}}

\newcommand{\ub}{{\bar u}}
\newcommand{\db}{{\bar d}}
\newcommand{\nb}{{\bar \n}}
\newcommand{\jb}{{\bar \j}}
\newcommand{\bx}{{\mathbf x}}
\newcommand{\bp}{{\mathbf p}}
\newcommand{\bk}{{\mathbf k}}
\newcommand{\bq}{{\mathbf q}}
\newcommand{\bl}{{\mathbf l}}
\newcommand{\pu}{{\mathbf p}_u}
\newcommand{\pd}{{\mathbf p}_d}
\newcommand{\pe}{{\mathbf p}_e}
\newcommand{\pv}{{\mathbf p}_\n}

\begin{document}

\markboth{Huang, $et.al.$}{Neutrino Emission From Pion Condensed
Quark Matter}

\catchline{}{}{}{}{}

\title{NEUTRINO EMISSION IN INHOMOGENEOUS PION CONDENSED QUARK MATTER}

\author{\footnotesize Xuguang Huang$^1$\footnote{e-mail: huangxg03@mails.tsinghua.edu.cn},
Qun Wang$^2$\footnote{e-mail: qunwang@ustc.edu.cn}, Pengfei
Zhuang$^1$\footnote{e-mail: zhuangpf@mail.tsinghua.edu.cn}}

\address{$^1$Physics Department, Tsinghua University,
Beijing 100084, China\\
$^2$Department of Modern Physics, University
of Science and Technology of China, Hefei, Anhui 230026, China}

\maketitle

\begin{history}
\received{(received date)}
\revised{(revised date)}
\end{history}

\begin{abstract}
It is believed that quark matter can exist in neutron star interior
if the baryon density is high enough. When there is a large isospin
density, quark matter could be in a pion condensed phase. We compute
neutrino emission from direct Urca processes in such a phase,
particularly in the inhomogeneous Larkin-Ovchinnikov-Fulde-Ferrell
(LOFF) states. The neutrino emissivity and specific heat are
obtained, from which the cooling rate is estimated.
\end{abstract}

\section{Introduction}
\label{introduction}

\begin{figure}[!htb]
\begin{center}
\includegraphics[width=6cm]{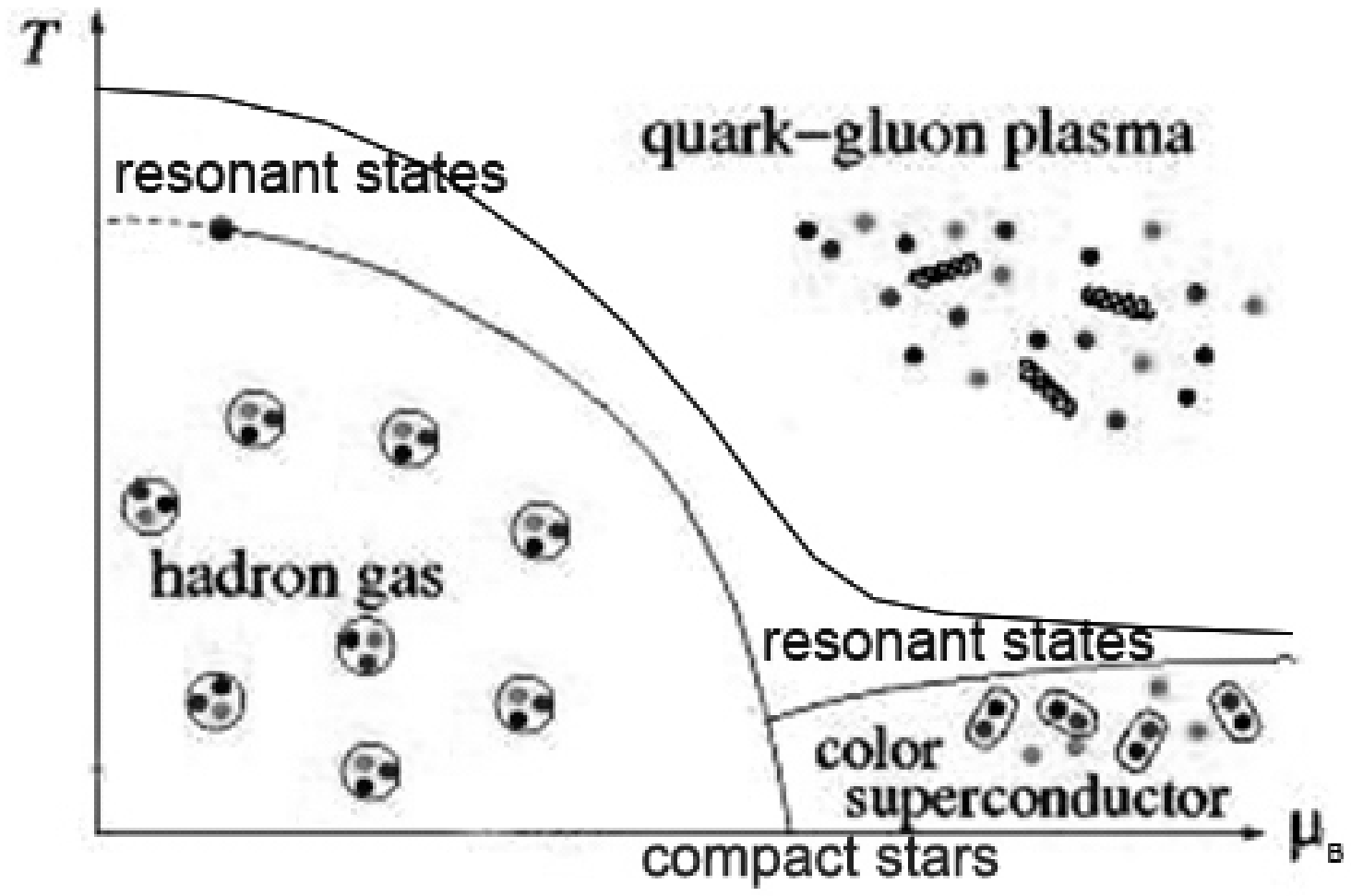}
\includegraphics[width=6cm]{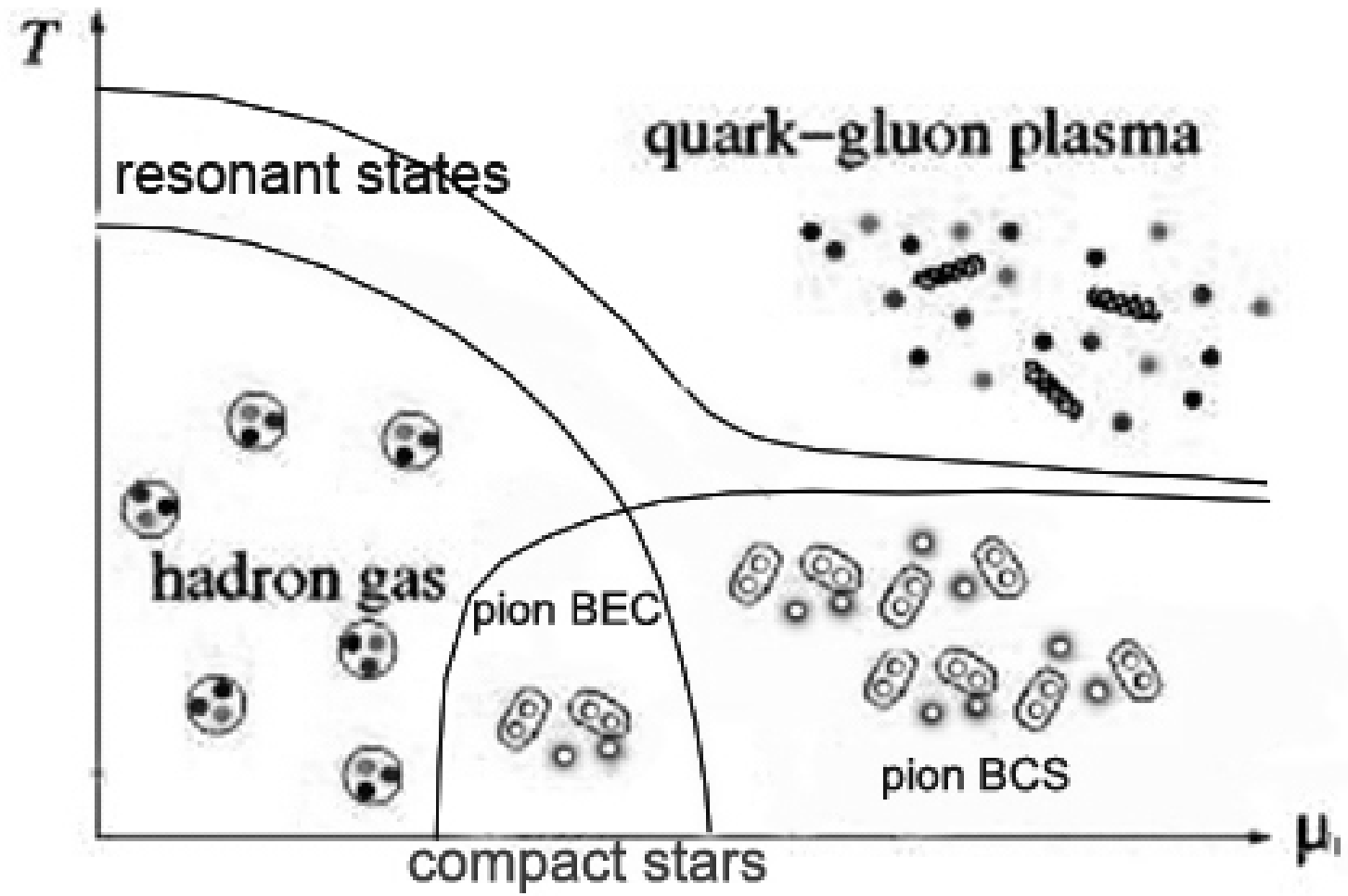}
\caption{The schematic phase diagrams of QCD on $T-\m_B$ and
$T-\m_I$ planes.} \label{phase}
\end{center}
\end{figure}

The phase structure of quantum chromodynamics (QCD) is one of the
most challenging problem in particle and nuclear physics. We
schematically illustrate in Fig.\ \ref{phase} the phase diagrams in
$T-\m_B$ and $T-\m_I$ plots, where $T$, $\m_B$ and $\m_I$ are the
temperature, baryon and isospin (or equivalently electron) chemical
potentials respectively. In the $T-\m_B$ diagram, the left panel of
Fig.\ \ref{phase}, the hadronic phase locates at low $T$ and low
$\m_B$ region and undergoes a phase transition or a crossover to the
deconfined quark phase at certain critical temperature $T_c$ or
baryon chemical potential $\m_{Bc}$ of the orders of $T_c\sim 200$
MeV or $\m_{Bc}\sim 1$ GeV. At very high temperature, the
quark-gluon-plasma (QGP), made of free quarks and gluons, forms. At
asymptotically high $\m_B$ but low $T$, the ground state of QCD is
the color-flavor-locked (CFL) superconductor \cite{alford1999} where
the condensation of quark pairs spontaneously breaks color and
chiral symmetries. At intermediate $T$ and $\m_B$, although quarks
and gluons are deconfined they are still strongly coupled. In this
regime, many QCD phases are proposed in recent years, such as, at
low $T$, two-flavor color superconductivity (2SC)
\cite{Ruester:2004eg}, gapless 2SC (g2SC) \cite{shovkovy2003},
gapless CFL (gCFL)\cite{alford2004}, spin-1 color superconductor
\cite{iwasaki1995,schafer2000,schmitt2002}, kaon condensation in the
CFL phase \cite{kaon-condensation}, et al.. For reviews of color
superconductivity, see, e.g. Ref. \cite{csc-reviews}. There may
exist resonance states at intermediate $T$, such as strongly coupled
QGP (sQGP) \cite{shuryak2006} at low $\m_B$ or the pseudo-gap phase
at moderate $\m_B$ \cite{kitazawa2005}. In the $T-\m_I$ diagram, the
right panel of Fig.\ \ref{phase}, the hadron phase is in the region
with low $T$ and low $\m_I$ while the QGP phase locates at very high
$T$. At low $T$, when $\m_I$ grows above the value of the pion mass
$m_\p$, the ground state turns out to be a Bose-Einstein
condensation (BEC) of pions, as $\m_I$ increases further and is
larger than about 230 MeV, the pion BEC crossover smoothly into the
BCS superfluid of quark-anti-quark pairs with the condensate
$\langle \ub i\g_5 d\rangle$ or $\langle \db i\g_5u\rangle$
\cite{son2001}. At intermediate $T$, resonance states such as sQGP
may occur at low $\m_I$ while the states with strong fluctuations of
thermally excited mesons or Cooper pairs are possible at moderate
$\m_I$.

In this paper we consider the quark matter cores in neutron stars.
The neutrino emission from direct Urca processes $d\ra u+e^-+\nb,
u+e^-\ra d+\n$ is the most efficient way of cooling in quark matter.
Sch\"afer and Schwenzer summarized the neutrino emissivities and
specific heats for a variety of color superconducting phases of
quark matter \cite{schafer2004}. Due to beta equilibrium the isospin
chemical potential $\m _I$ is nonzero, quark matter could be a pion
BEC ($\m_I<230$ MeV) or a BCS superfluid ($\m_I > 230$ MeV) when $\m
_I > m_\p$ \cite{son2001}. On the other hand, the baryon chemical
potential is still large (in the order of 1 GeV) and makes a big
mismatch between the fermi surfaces of the pairing quarks $u(\bar
u)$ and $\bar d(d)$, thus the BEC or BCS state is gapless. Such a
gapless phase is stable in the BEC region but unstable in the BCS
one with respect to the formation of nonzero LOFF momentum. In a
previous work \cite{huang2007}, we have studied the neutrino
emissivity and cooling rate due to Urca processes for the gapless
pion condensed quark matter in the BEC region. In this paper we
study the neutrino emission in the LOFF phase. We work in two-flavor
case in the moderate baryon density, where the role of strange
quarks is not important.

Our units are $\hbar=k_B=c=1$ except particular specifications.
As a convention, we denote a 4-momentum as $K^\mu =(k_0,\mathbf{k})$, and
its 3-momentum magnitude as $k=|\mathbf{k}|$.

\section {Quark Propagator\label{quark}}

Our starting point is the two flavor Nambu-Jona-Lasinio Lagrangian of QCD
\begin{eqnarray}
\label{NJL}
\cl=\jb(i\g^\m\pt_\m+\m\g_0-m_0)\j+g[(\jb\j)^2+(\jb
i\vec{\t}\g_5\j)^2],
\end{eqnarray}
where $\j=(u,d)^{\rm T}$ is the quark fields, $g$ is the coupling
constant and $\vec{\t}$ is the Pauli matrices. We have introduced
the chemical potential matrix in flavor space,
$\m=\mathrm{diag}(\m_u, \m_d)=(\m+\d\m, \m-\d\m)=(\m_B/3+\m_I/2,
\m_B/3-\m_I/2)$ with $\m_B, \m_I$ the baryon and isospin chemical
potential respectively. We assume that the $\b$-equilibrium is
reached $\m_d=\m_u+\m_e$, which gives $\m_I=-\m_e$. In the chiral
limit and without the chemical potentials, the Lagrangian
(\ref{NJL}) respects the symmetry $U_B(1)\otimes SU_V(2)\otimes
SU_A(2)$ corresponding to the baryon number, isospin vector and
pseudovector conservation respectively. However, the presence of the
chemical potentials explicitly break the isospin symmetry down to
$U_V(1)$ with the conserved quantum number of $\t_3$, and chiral
symmetry down to $U_A(1)$ with the conserved quantum number of
$i\g_5\t_3$. By introducing the chiral condensate
$\s=-2g\lan\jb\j\ran$ and pion condensates
$\p^-=\D e^{-2i\bl\cdot\bx}=4g\langle\ub i\g_5 d\ran$,
$\pi^+=\D e^{2i\bl\cdot\bx}=4g\langle\db i\g_5 u\ran$,
we arrive at the mean field Lagrangian
\begin{eqnarray}
\cl_\mf=\jb\lb\begin{array}{cc}i\g^\m \pt_\m+\m_u\g_0-m & \p^+i\g_5
\\ \p^-i\g_5 & i\g^\m \pt_\m+\m_d\g_0-m
\end{array}\rb\j
-\frac{\s^2+\D^2}{4g},
\end{eqnarray}
with the effective quark mass $m=m_0+\s$. There are mismatches in
Fermi surfaces of anti-u and d quarks or anti-d and u quarks by the
baryon chemical potential, hence pion condensates with nonzero total
momentum or the LOFF states may be favored. The formation of the
condensate $\s\sim\lan\jb\j\ran$ breaks the $U_A(1)$ chiral symmetry
spontaneously with the Goldstone boson $\p_0\sim\jb i\g_5\t_3\j$,
and that of pion condensates $\D\sim \langle\db i\g_5u\ran\sim
\langle\ub i\g_5d\ran$ break the $U_V(1)$ isospin symmetry
spontaneously. The translational and rotational symmetries are
spontaneously broken by nonzero LOFF momentum $\bl$. The partition
function of the system can be written as a functional integral
\begin{equation}
Z=\int [d\jb ] [d\j ] \exp{\left( \int_0^\b d\t\int
d^3\bx\cl_\mf\right) } .
\end{equation}
Rewriting the quark fields via a gauge transformation which leaves the
partition function unchanged, $\c_u(x)=u(x)e^{-i\bl\cdot\bx}$,
$\c_d(x)= d(x)e^{i\bl\cdot\bx}$ (we still call $\c _{u,d}$ quark
fields), the inverse quark propagator in flavor and momentum space reads
\begin{eqnarray}
\label{s-inverse}
S^{-1}(K)=\lb\begin{array}{cc}\g^\m
K_\m-\bl\cdot\g+\m_u\g_0-m & \D i\g_5 \\
\D i\g_5 & \g^\m K_\m+\bl\cdot\g+\m_d\g_0-m
\end{array}\rb.
\end{eqnarray}
Note that $K^\m$ represents the 4-momentum of the $\c$ fields
instead of the $\j$ fields. The propagator is written as
\begin{eqnarray}
\label{propagator}
S(K)=\lb\begin{array}{cc} S_{uu}(K) & S_{ud}(K) \\
S_{du}(K) & S_{dd}(K)
\end{array}\rb.
\end{eqnarray}
A straightforward calculation from Eq. (\ref{s-inverse})
gives the four elements
\begin{eqnarray}
S_{uu}(K)&=&\frac{(\g^\m K_{+\m}+m)(K_-^2-m^2-\D^2)+2\D^2\g^\m
l_\m}{(K_+^2-m^2+\D^2)(K_-^2-m^2+\D^2)-\D^2[(K_++K_-)^2-4m^2]},\non
S_{dd}(K)&=&\frac{(\g^\m K_{-\m}+m)(K_+^2-m^2-\D^2)-2\D^2\g^\m
l_\m}{(K_+^2-m^2+\D^2)(K_-^2-m^2+\D^2)-\D^2[(K_++K_-)^2-4m^2]},\non
S_{ud}(K)&=&\frac{(\g^\m K_{+\m}+m)(K_-^2-m^2-\D^2)+2\D^2\g^\m
l_\m}{(K_+^2-m^2+\D^2)(K_-^2-m^2+\D^2)-\D^2[(K_++K_-)^2-4m^2]}\non
&&\times \frac{\g^\m K_{-\m}-m}{K_-^2-m^2}i\g^5\D,\non
S_{du}(K)&=&\frac{(\g^\m K_{-\m}+m)(K_+^2-m^2-\D^2)-2\D^2\g^\m
l_\m}{(K_+^2-m^2+\D^2)(K_-^2-m^2+\D^2)-\D^2[(K_++K_-)^2-4m^2]}\non
&&\times \frac{\g^\m K_{+\m}-m}{K_+^2-m^2}i\g^5\D,\non
\end{eqnarray}
where $K_\pm^\m=(k_0+\m\pm\d\m,\bk\pm\bl)$ and $l^\m=(\d\m, \bl)$.
The excitation spectra of the quasi-particles can be obtained by
solving the equation $\det S^{-1}(k_0,\bk)=0$ or equivalently the
roots of the denominator of the propagator for $k_0$,
\begin{eqnarray}
0&=&(K_+^2-m^2+\D^2)(K_-^2-m^2+\D^2)-\D^2[(K_++K_-)^2-4m^2]\non
&\approx&[(k_0+\m+\ve^-_{\bk,\bl})^2-(\ve_{\bk,\bl}^+ +\d\m)^2
-\D^2]\non
&&\times [(k_0+\m-\ve_{\bk,\bl}^-)^2-(\ve_{\bk,\bl}^+-\d\m)^2-\D^2],
\end{eqnarray}
where $\ve_{\bk,\bl}^\pm=(E_{\bk+\bl}\pm E_{\bk-\bl})/2$ with
$E_\bk\equiv\sqrt{\bk^2+m^2}$. To arrive at the last line, we have
taken the assumption that both $\D$ and $\bl$ are small comparing to
the quark Fermi momenta in the LOFF phase. We have four excitation
branches, $E_r^a(\bk,\bl)=-r\sqrt{(\ve_{\bk,\bl}^+ -
a\d\m)^2+\D^2}-(\m - a\ve_{\bk,\bl}^-)$, with $a,r=\pm$. Taking the
same approximation to the numerators in the elements of the quark
propagator, we neglect the terms proportional to $\D^2\bl$.
Thus we rewrite the elements of the quark propagator as,
\begin{eqnarray}
\label{suu-sdd}
S_{uu}(K)&\simeq&\sum_{a,r=\pm}\frac{B_{r}^{a}(\bk ,\bl )\L^{a}_{\bk+\bl }\g_0}{k_0-E_r^a(\bk,\bl)},\non
S_{dd}(K)&\simeq&\sum_{a,r=\pm}\frac{B_{-r}^{a}(\bk ,\bl )\L^{-a}_{\bk -\bl }\g_0}{k_0-E_r^a(\bk,\bl)},
\end{eqnarray}
where we have introduced the energy projectors,
\begin{equation}
\L^a_{\bk }=\frac{1}{2}\left[1+a\frac{\g_0(\g\cdot\bk+m)}{E_\bk}\right],
\end{equation}
and the Bogoliubov coefficients,
\begin{equation}
\label{bog-coeff}
B^{a}_r(\bk,\bl)=\frac 12 \left[1-ar\frac{\varepsilon ^+_{\bk,\bl}-a\delta\mu}
{\sqrt{(\varepsilon ^+_{\bk,\bl}-a\delta\mu)^2+\Delta^2 }}\right],
\end{equation}

\section{Neutrino Emissivity}
\label{neutrino}

Since the characteristic energy scale of the Urca processes is much
lower than the W-boson mass, we can use the Fermi
current-current interaction to describe the Urca processes,
\begin{equation}
\label{c-current}
\cl_{\rm int} = \frac{G}{\sqrt{2}}J^{\m}J_{\m}^{\dag},
\end{equation}
where the weak currents are defined by
\begin{eqnarray}
J^{\m}(x)&=&\nb\g^\m(1-\g_5)e
+e^{-2i\bl\cdot\bx}\bar{\c}_u\g^\m(1-\g_5)\c_d, \non
J_{\m}^\dag(x)&=&\bar{e}\g_\m(1-\g_5)\n+e^{2i\bl\cdot\bx}\bar{\c}_d
\g_\m(1-\g_5) \c_u ,
\label{eq:jmu}
\end{eqnarray}
where $i$ and $\bar{i}$ for $i=e,\n$ denote spinor fields for
electrons and neutrinos. Here
$G=G_F\cos\theta_C\approx1.13488\times10^{-11}{\rm MeV}^{-2}$ is the
four-fermion coupling constant.

In the $\b$-equilibrated quark matter, the neutrino emissivity,
defined as the total energy per unit time and per unit volume
carried away by neutrinos and anti-neutrinos in escaping a neutron
star can be written as \cite{huang2007},
\begin{eqnarray}
\label{emss}
\e &=&  2G^2 \int\frac{d^3\pe}{(2\p)^32E_e}
\int\frac{d^3\pv}{(2\p)^3 2E_{\n}} E_{\n} n_B(-E_{e}+\m_e+E_{\n})
n_F(E_{e}-\m_e) \non && \times L^{\l\s}(P_\n,P_e)\im\P^R_{\l\s}
(E_{e}-\m_e-E_{\n},\pe-\pv),
\end{eqnarray}
Here the on-shell 4-momenta for leptons is denoted by
$P_i=(E_i,\bp_i)$ ($i=e,\n$), where the energies are $E_i \equiv
E_{p_i}=\sqrt{p_i^2+m_i^2}$ with $m_{\n}=0$ and $m_{e}\approx 0$.
The factor $2$ in the front is due to the approximately identical
cross sections for $\b$-decay and electron capture processes at very
low temperatures far from the condensation-normal transition
temperature $T_c$. But as the temperature approaches $T_c$, one
cannot take the contributions from the two processes as equal, as is
demonstrated in Ref. \cite{huang2007}. We used
$n_B(x)=(e^{x/T}-1)^{-1}$ and $n_F(x)=(e^{x/T}+1)^{-1}$ to denote
the Bose-Einstein and Fermi-Dirac distribution functions, which
maintain the detailed balance and Pauli enhancing/blocking effects
for electrons in the electron capture/$\b$-decay processes. Note
that there are two identical terms in evaluating the emissivity
(\ref{emss}), see Fig. 3 of Ref. \cite{Wang:2006xfa}, which gives
the correct overall factor and can easily be overlooked. We have
dropped the chemical potential of neutrinos since there is no
accumulations of neutrinos at typical temperatures inside an aging
star. The leptonic tensor reads
\begin{eqnarray}
L^{\l\s}(P_\nu,P_e) &=&\Tr[\g^\l(1-\g_5)\g\cdot
P_e\g^\s(1-\g_5)\g\cdot P_\n]\non &=& 8 [P_e^\l P_\n^\s+P_\n^\l
P_e^\s-P_e\cdot P_\n g^{\l\s}-i\e^{\l\a\s\b} P_{e\a}P_{\n\b}].
\end{eqnarray}
The W-boson polarization tensor $\P^{\l\s}(q_0,\bq)$ can be written as
\begin{eqnarray}
\label{pi} \P^{\l\s}(q_0,\bq) & = & N_c T \sum
_n\int\frac{d^3\pu}{(2\p )^3}\Tr [\g^\l (1-\g_5)
S_{uu}(p_{u0},\pu)\g^\s (1-\g_5) S_{dd}(p_{d0},\pd)] \non &=&N_c\sum
_{a,b,r,s}\int\frac{d^3\pu}{(2\p)^34E_uE_d}
\frac{n_F[E_r^a(\pu,\bl)]-n_F[E_s^{-b}(\pd,\bl)]}
{q_0-E_s^{-b}(\pd,\bl)+E_r^a(\pu,\bl)}\non &&\times B_r^{a}(\pu,\bl)
B^{-b}_{-s}(\pd,\bl) H^{\l\s}_{ab}(P_u,P_d),
\end{eqnarray}
where $p_{d0}= p_{u0}+q_0$, $\pd=\pu+\bq+2\bl$,
$P_u=(E_u,a\pu+a\bl)$ and $P_d=(E_d,b\pd-b\bl)$ with
$E_u=E_{\pu+\bl},E_d=E_{\pd-\bl}$. Here $\pu,\pd$ are the momenta of
$\c_u,\c_d$. The bosonic and fermionic Matsubara frequencies are
given by $q_0=i 2m\p T$ and $p_{u0}=i(2n+1)\p T$ ($m,n$ are
integers). The quark tensor $H^{\l\s}_{ab}(P_u,P_d)$ is defined by
\begin{eqnarray}
H^{\l\s}_{ab}(P_u,P_d) & = & 4E_uE_d \Tr
[\g^\l(1-\g_5)\L^{a}_{\pu+\bl}\g_0
\g^\s(1-\g_5)\L^{b}_{\pd-\bl}\g_0]\non &=& 8[P_u^\l P_d^\s+P_d^\l
P_u^\s-P_u \cdot P_d g^{\l\s}-i\e^{\l\a\s\b}P_{u\a}P_{d\b}].
\end{eqnarray}
By an analytic extension for the Matsubara frequency $q_0=i 2m\pi
T\ra q_0+i0^+$, where the second $q_0$ is real, the imaginary part
of the retarded polarization tensor of W-bosons can be read out
directly,
\begin{eqnarray}
\label{impi}
\im\P^{\l\s}_R(q_0,\bq) & = & \p N_c\sum _{a,b,r,s}
\int\frac{d^3\pu}{(2\p )^3 4E_uE_d}
\d[q_0-E_s^{-b}(\pd,\bl)+E_r^a(\pu,\bl)] H_{ab}^{\l\s}(P_u,P_d)\non
&&\times\frac{n_F[E_r^a(\pu,\bl)]n_F[-E_s^{-b}(\pd,\bl)]}{n_B(-q_0)}
B_r^{a}(\pu,\bl) B^{-b}_{-s}(\pd,\bl) .
\end{eqnarray}
Substituting Eq.(\ref{impi}) into Eq.(\ref{emss}) we arrive at
\begin{eqnarray}
\e & = & 4N_c \sum _{a,b,r,s}\int\frac{d^3\pe}{(2\p)^3 2E_e}
\frac{d^3\pv}{(2\p)^3 2E_\n}\frac{d^3\pu}{(2\p)^32 E_u}
\frac{d^3\pd}{(2\p)^3 2E_d} \non &&\times E_\n (2\p)^4
\d[E_e-\m_e-E_{\n}+E_r^a(\pu,\bl) - E_s^{-b}(\pd,\bl)] \non &&
\times \d^3(\pe-\pv+\pu-\pd+2\bl) B_r^{a}(\pu,\bl)
B^{-b}_{-s}(\pd,\bl)\non && \times n_F(E_e-\mu_e)
n_F[E_r^a(\pu,\bl)]n_F[-E_s^{-b}(\pd,\bl)] |M_{ab}|^2. \label{e2}
\end{eqnarray}
We have introduced the shorthand notation $|M_{ab}|^2$, the
spin-averaged scattering matrix element of $\beta$-decay or electron
capture \cite{iwamoto1980},
\begin{equation}
\label{amplitude} |M_{ab}|^2 = \frac{G^2}{4}
L_{\l\s}(P_e,P_\n)H_{ab}^{\l\s}(P_u,P_d) =64 G^2(P_e\cdot
P_u)(P_\n\cdot P_d),
\end{equation}
where the energy projection indices $a,b$ are hidden in the quark momenta.

In the LOFF phase, chiral symmetry is almost restored, so we can
safely set quark masses zero, $m_{u,d}=0$. We consider the low temperature and
high chemical potentials for quarks, the dominant
contribution of the phase space integral in quark momenta
comes from the gapless modes of the excitations with
positive energies, i.e. terms with $r=s=-$.
Also, since the gap parameter and the LOFF momentum are
small comparing to the quark chemical potentials, the Bogoliubov
coefficients with $a=-$ and $b=-$ are strongly suppressed,
so we only keep the term with $a=+$ and $b=+$.
For electrons, the relevant momenta are near the chemical potential.
Recalling that $E_-^+(\pu,\bl)$ and $E_-^-(\pd,\bl)$
are the dispersion relations for quasi-u-quarks and quasi-d-quarks,
one can obtain the gapless momenta as
\begin{equation}
p_{u/d}^0\approx (1-\kappa )[\sqrt{(\m \mp l\cos\h_{u/d})^2-\D^2}\pm \d\m]
\end{equation}
for $u/d$ quarks, where $\kappa$ is due to the Fermi-liquid correction.
In perturbative QCD we have $\k=2\a_s/(3\p)$ with $\a_s$
being the strong coupling constant \cite{baym1976}.
In NJL model we have \cite{wang2006} $\k=4g\m_B^2/(3\p^2)$.
We have set the z-direction along the LOFF momentum
$\bl$ and denote as $\h_i$ and $\f_i$ the polar and azimuthal angles
of $\bp_i,i=e,\n,u,d$. Near the gapless momenta, we can make expansion
$E_-^+(\pu,\bl)\approx v_u|p_u-p_u^0|$ and
$E_-^-(\pd,\bl)\approx v_d|p_d-p_d^0|$ with velocities
$v_{u/d}=\sqrt{1-\D^2/(\m\mp l\cos\h_{u/d})^2}$.
The matrix element is now evaluated as
\begin{eqnarray}
|M_{++}|^2 &\approx & 64 G^2 p_ep_\n E_uE_d [1-(1-\kappa
)\cos\h_{ue}] [1-(1-\kappa )\cos\h_{d\n}],
\end{eqnarray}
where $\h_{d\n }$ and $\h_{ue}$ are angles between $\pd -\bl$ and
$\pv$ and between $\pu +\bl $ and $\pe$. Since both $\D,\bl$ are
small and $p_v\sim T\ll\m,\d\m$ is negligible, the delta function
for the energies in Eq. (\ref{e2}) can be rewritten in the form
$\d[E_e-\m_e-E_{\n}+E_-^+(\pu,\bl) - E_-^-(\pd,\bl)]\approx
\m_e/(p_u^0p_d^0)\d(\cos\h_{ud}-\cos\h_{ud}^0)$ with $\h_{ud}$ the
angle between $\pd-\bl$ and $\pu+\bl$ and
$\cos\h_{ud}^0\approx[(p_u^{0})^2+(p_d^{0})^2-2\m_e
l(\cos\h_u+\cos\h_d)]/(2p_u^0p_d^0)$. The integration over electron
three-momentum can be carried out by the delta function about the
momentum conservation. Now we get,
\begin{eqnarray}
\e&\approx&\frac{914}{315} \p^7N_cG^2 \m_e T^6\int
\frac{d\O_\n}{(2\p)^3}\frac{d\O_u}{(2\p)^3}\frac{d\O_d}{(2\p)^3}
\frac{p_u^0p_d^0}{v_uv_d} B_-^+(p_u^0)B_+^-(p_d^0) \non &&\times
\d(\cos\h_{ud}-\cos\h_{ud}^0) [1-(1-\kappa)\cos\h_{ue}][1-(1-\kappa
)\cos\h_{d\n}] \non &=&\frac{457}{315} \p^5N_cG^2 \m_e
T^6\int\frac{d\O_u}{(2\p)^3}\frac{d\O_d}{(2\p)^3}
\frac{p_u^0p_d^0}{v_uv_d} B_-^+(p_u^0)B_+^-(p_d^0) \non && \times
\d(\cos\h_{ud}-\cos\h_{ud}^0)[1-(1-\kappa)\cos\h_{ue}], \label{e3}
\end{eqnarray}
where we have used \cite{morel1962} $\int_0^\infty dp_\n dp_u
dp_dp_\n^3
n_F(p_\n+|p_d-p_d^0|-|p_u-p_u^0|)n_F(|p_u-p_u^0|)n_F(-|p_d-p_d^0|)
\approx 457\p^6T^6/5040 $. By setting $\D=l=0$ in \eq{e3}, one can
recover the well-known result for the neutrino emissivity in normal
quark matter \cite{iwamoto1980},
\begin{equation}
\e_0\approx\frac{457}{2520} \p N_cG^2 \m_e\m_u\m_d
\lb1+\frac{\m_d}{\m_u}\rb\k T^6. \label{e0}
\end{equation}

\begin{figure}[!htb]
\begin{center}
\includegraphics[width=7cm]{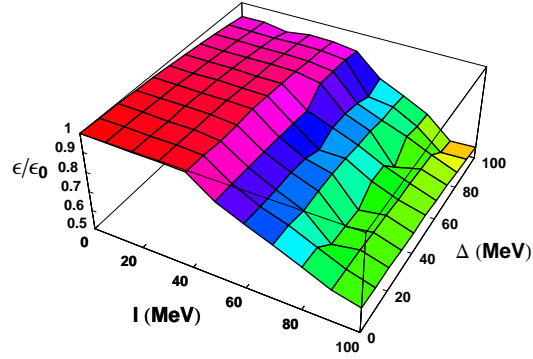}
\caption{(Color online) The neutrino emissivity in a LOFF pion
superfluid as functions of the pion condensate $\D$ and the
LOFF momentum $l$. The parameters are set to
$\k=2/(3\p)$, $\m=300$ MeV, $-\d\m=125$ MeV.} \label{edl}
\end{center}
\end{figure}

We make numerical evaluation of the neutrino emissivity.
We choose $\k=2/(3\p)$, $\m=300$ MeV, $-\d\m=125$ MeV.
These are typical values to support the possible LOFF pion
superfluid. In Fig. \ref{edl}, the neutrino emissivity
in unit of $\e_0$ is presented as function of the pion condensate
$\D$ and the LOFF momentum $l$. Due to the
gapless nature, along the $\D$ axis, these is no exponential
but an approximate quadratic suppression. The appearance of the LOFF
momentum also slightly lowers the neutrino emissivity.
Note that the approximation we made here is somewhat
different from that we did in the previous work \cite{huang2007}.
In this paper we evaluate the Bogoliubov coefficients at
gapless momenta and result in a suppression in the
emissivity. In the previous work the Bogoliubov coefficients
are taken to be 1 approximately which leads in turn to
an emissivity enhancement. The latter approximation
is valid even at high temperatures,
but the former one should be adopted only at low temperatures.

\section{Cooling Rates}
\label{cooling}
To get the cooling behavior, one must know the specific heat
of the pion superfluid which can be derived from its definition
$c_V(T)=T({\pt S}/{\pt T})_V$ with $S$ being the entropy density,
\begin{eqnarray}
S&=&-2N_c\sum_{r,a}\int\frac{d^3\bk}{(2\p)^3}
\Big\{n_F[E_r^a(\bk,\bl)] \ln n_F[E_r^a(\bk,\bl)] \non
&&+n_F[-E_r^a(\bk,\bl)]\ln n_F[-E_r^a(\bk,\bl)]\Big\},
\end{eqnarray}
where the pre-factor $2N_c$ comes from the degeneracies of spins and
colors. Since we work at low temperatures and high quark chemical
potentials, the dominant contribution to the specific heat are from
the gapless momenta of positive energy excitations,
\begin{eqnarray}
c_V(T) &\approx& 2N_c\sum_a\int
\frac{d^3\bk}{(2\p)^3}n_F[E_-^a(\bk,\bl)]n_F[-E_-^a(\bk,\bl)]
\frac{[E_-^a(\bk,\bl)]^2}{T^2}\non
&\approx&\frac{2\p^2N_c}{3}T\sum_i\int
\frac{d\O}{(2\p)^3}\frac{(p_i^0)^2}{v_i}, \label{cv1}
\end{eqnarray}
where we have used the fact that at low temperatures $\D,\bl$ are
almost constants. When $\D$ and $l$ vanish, we reproduce the
specific heat of normal two-flavor quark matter $c_{V0}=\gamma T$
with $\gamma=2N_c(\m_u^2+\m_d^2)T/6$.

The time evolution of the temperature can be obtained by solving the
following equation $t-t_0=-\int_{T_0}^TdT'c_V(T')/\e(T')$, where
$T_0$ is the temperature at an initial time $t_0$. Substituting
\eq{e3} and \eq{cv1} into this equation, we arrive at
\begin{eqnarray}
\label{rate2} t-t_0=-\frac{210}{457\p^3G^2\m_e}\int_{T_0}^T
\frac{d\, T'}{T'^5} \frac{G(\D,l)}{F(\D,l)},
\end{eqnarray}
where we have introduced the notations
\begin{eqnarray}
F(\D,l)&=& \int\frac{d\O_u}{(2\p)^3}\frac{d\O_d}{(2\p)^3}
\frac{p_u^0p_d^0}{v_uv_d} B_-^+(p_u^0)B_+^-(p_d^0)
\d(\cos\h_{ud}-\cos\h_{ud}^0)[1-(1-\kappa)\cos\h_{ue}] \non
G(\D,l)&=&\sum_{i=u,d}\int\frac{d\O}{(2\p)^3}\frac{(p_i^0)^2}{v_i}.
\end{eqnarray}

\begin{figure}[!htb]
\begin{center}
\includegraphics[width=7cm]{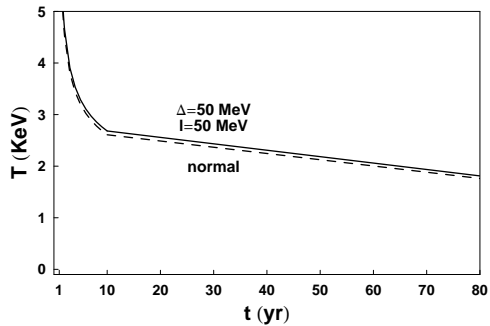}
\caption{The cooling curves of the pion superfluid in the LOFF phase
and of normal quark matter with $T_0=0.1$ MeV at $t_0=1$ yr.
The parameters are the same as in Fig. \ref{edl}. }
\label{Tt}
\end{center}
\end{figure}

Fig. \ref{Tt} shows the time evolution of the pion superfluid in the
LOFF phase with the initial temperature $T_0=0.1$ MeV at $t_0=1$ yr.
For comparison, we also show the curve of normal quark matter. We
choose $\D=l=50$ MeV. The other parameters are chosen as the same as
in Fig. \ref{edl}. We see that the cooling of the LOFF pion
superfluid is slightly slower than that for normal quark matter.
This situation is very similar to the case of inhomogeneous CFL
quark matter, see\cite{anglani2006}.

\section {Summary}
\label{summary} If the isospin (or equivalently electron) chemical
potential in a neutron star is large enough, the LOFF pion
superfluid can be a possible ground state in the interior matter of
the neutron star. We calculated the neutrino emissivity, specific
heat and then the cooling rate through direct Urca processes in this
phase. The main results are shown in \eq{e3} for the neutrino
emissivity and in \eq{rate2} for the cooling rate. At low
temperatures, the neutrino emissivity is slightly smaller in the
LOFF pion superfluid than that in normal quark matter, as indicated
in Fig. \ref{edl} which leads to a slightly slower cooling rate for
the LOFF phase as shown in Fig. \ref{Tt}.

\vspace{1cm}

{\bf Acknowledgments:} Q.W. is supported in part by the startup
grant from University of Science and Technology of China (USTC) in
association with 'Bai Ren' project of Chinese Academy of Sciences
(CAS) and by National Natural Science Foundation of China (NSFC)
under the grant 10675109.

\end{document}